\documentclass[runningheads]{llncs}
\usepackage{graphicx}
\usepackage{multirow}

\begin{document}
\title{Colonoscopy Polyp Detection: \\Domain Adaptation From Medical Report Images to Real-time Videos}

\author{Zhi-Qin Zhan\inst{1} \and
Huazhu Fu\inst{2} \and
Yan-Yao Yang\inst{1,3} \and
Jingjing Chen\inst{1} \and
Jie Liu \inst{1,3} \and
Yu-Gang Jiang\inst{1}}
\authorrunning{Z. Zhan et al.}
%
\institute{Fudan University\\
\email{\{zqzhan14, 12301010048, chenjingjing, jieliu, ygj\}@fudan.edu.cn} \and
Inception Institute of Artificial Intelligence\\
\email{hzfu@ieee.org} \and
Department of Digestive Diseases, Huashan Hospital}
\maketitle              
\begin{abstract}

Automatic colorectal polyp detection in colonoscopy video is a fundamental task, which has received a lot of attention.
Manually annotating polyp region in a large scale video dataset is time-consuming and expensive, which limits the development of deep learning techniques.
A compromise is to train the target model by using labeled images and infer on colonoscopy videos.
However, there are several issues between the image-based training and video-based inference, including domain differences, lack of positive samples, and temporal smoothness.
To address these issues, we propose an \textbf{I}mage-\textbf{v}ideo-joint pol\textbf{y}p detection network (Ivy-Net) to address the domain gap between colonoscopy images from historical medical reports and real-time videos.
In our Ivy-Net, a modified mixup is utilized to generate training data by combining the positive images and negative video frames at the pixel level, which could learn the domain adaptive representations and augment the positive samples.
Simultaneously, a temporal coherence regularization (TCR) is proposed to introduce the smooth constraint on feature-level in adjacent frames and improve polyp detection by unlabeled colonoscopy videos.
For evaluation, a new large colonoscopy polyp dataset is collected, which contains 3056 images from historical medical reports of 889 positive patients and 7.5-hour videos of 69 patients (28 positive).
The experiments on the collected dataset demonstrate that our Ivy-Net achieves the state-of-the-art result on colonoscopy video.

\end{abstract}
\section{Introduction}
\begin{figure} [t]
	\includegraphics[width=\textwidth]{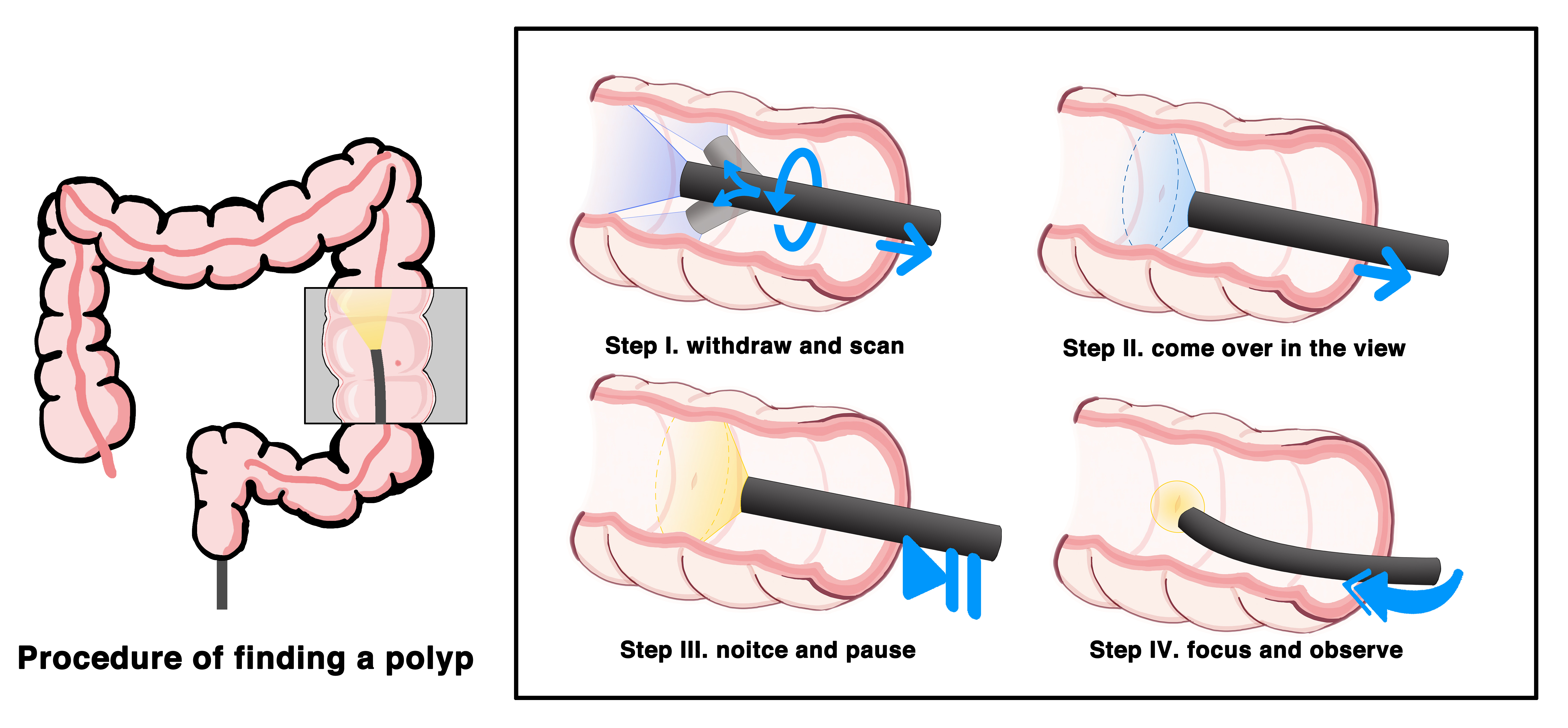}
	\caption{Procedure of Finding a Polyp.
		Modern colonoscopy inserts a long flexible tube from rectum to ileum.
		That tube contains artificial illuminants, cameras, and channels for air, water as well as devices.
		Endoscopists observe through cameras during withdrawal of the tube.
		They can bend and rotate the tube's terminal to adjust the visual field.
		A typical procedure of finding a polyp into four steps, and CADe is most helpful at Step II.}
	\label{fig.operation}
\end{figure}

Colonoscopy allows Colorectal Cancer (CRC), the third most common cancer globally, to be preventable \cite{brenner2018colorectal,arnold2017global}.
Before threatening health, CRC performs as colon polyps that could be detected and treated by colonoscopy at an early stage \cite{strum2016colorectal,early2012appropriate}.
Thus precise polyp detection is essential.
However, due to the variety of colon polyps in morphology, distinguishing them from normal structures is challenging.
There is an ineluctable polyp miss rate even for professional endoscopists \cite{leufkens2012factors,van2006polyp}.
Computer-aided Detection (CADe) of colon polyps is therefore in demand \cite{alagappan2018artificial}.
Moreover, different from other medical imaging tasks such as CT and MRI,
the observation of colonoscopy is closely related to endoscopists' operation.
Fig.~\ref{fig.operation} shows a typical procedure of finding a colon polyp, which could be summarized into four steps as:
Step I)~Scan the colon for polyps;
Step II)~A polyp comes into view;
Step III)~The endoscopist notices a potential polyp and stops withdrawing;
Step IV)~Adjust the visual field to have a closer observation, and maybe capture photos for the report.
Endoscopists need to judge and act at Step III, in order to observe closer and record at Step IV.
Hence, CADe is most clinically needed at Step II when polyps first come in view, i.e., in real-time video streams.

With the recent rise of deep learning, research on CADe of colon polyps has flourished,
both at image-level \cite{wang2019real,wang2020effect,de2019training} and video-level \cite{yu2016integrating,zhang2018polyp,qadir2019improving,itoh2019stable}.
Wang et al. demonstrated the clinical feasibility of CADe via further prospective trials \cite{wang2019real,wang2020effect},
with a CNN model trained with still colonoscopy images.
However, its video-level performance is not satisfactory \cite{mori2018detecting}.
Other research also found the similar phenomenon in \cite{urban2018deep,rees2019artificial}.
We attribute this performance drop to the semantic gap between Step IV training images and Step II testing video frames (see Fig.~\ref{fig.operation}).
If so, models trained directly with Step II videos work best theoretically.
But in clinical practice, such video data is much harder to obtain, especially for positive samples.
Carefully captured Step IV colonoscopy images are available from historical medical report database,
while videos need to be specially recorded from scratch.
Sufficient number of positive patients are difficult to find in a short time.
Lack of positive samples may cause the model to overfit small datasets and hard to generalize.
Besides, video annotation is much more costly than images.
It is difficult to collect a large scale video dataset with high quality annotation for deep learning techniques.
In summary, training model with images has the domain gap between the image and video, simultaneously, training model with videos will suffer from the limitation of video data size.
\textit{Thus, addressing the domain gap between colonoscopy images from historical medical reports and real-time videos is an effective solution.}

In this paper, we consider joint utilization of colonoscopy images and videos to alleviate the dilemma above. \textbf{First,} we defined an image-video-joint colonoscopy polyp detection task, which aims to train the model by using labeled images and negative video frames and infer on the target colonoscopy videos.
\textbf{Second,} we propose an \textit{\textbf{I}mage-\textbf{v}ideo-joint pol\textbf{y}p detection network} (Ivy-Net) to address the domain gap between colonoscopy images from historical medical reports and real-time videos.
In our Ivy-Net, a modified mixup is utilized to generate training data by randomly combining the images and negative video frames at the pixel level.
\textbf{Third,} we further exploit the temporal information of videos, by using temporal coherence regularization (TCR) on feature-level during model training.
\textbf{Fourth,} we collect a large colonoscopy dataset that combines 3056 images from historical medical reports of 889 positive patients (diagnosed as colon polyp) and 7.5-hour colonoscopy videos of 69 patients (28 positive).
All images are annotated with bounding boxes;
all videos are provided with time slice labels of positive or negative;
in total 5050 of positive video frames are annotated with bounding boxes.
\textbf{Moreover,} the experiments on our collected dataset demonstrate that our Ivy-Net achieves the state-of-the-art result, and improves the average precision (IoU = 0.5) of polyp detection on colonoscopy video from 0.631 to 0.717.
\textbf{All the codes and dataset will be released upon publication of this work.}


\begin{figure} [t]
	\includegraphics[width=\textwidth]{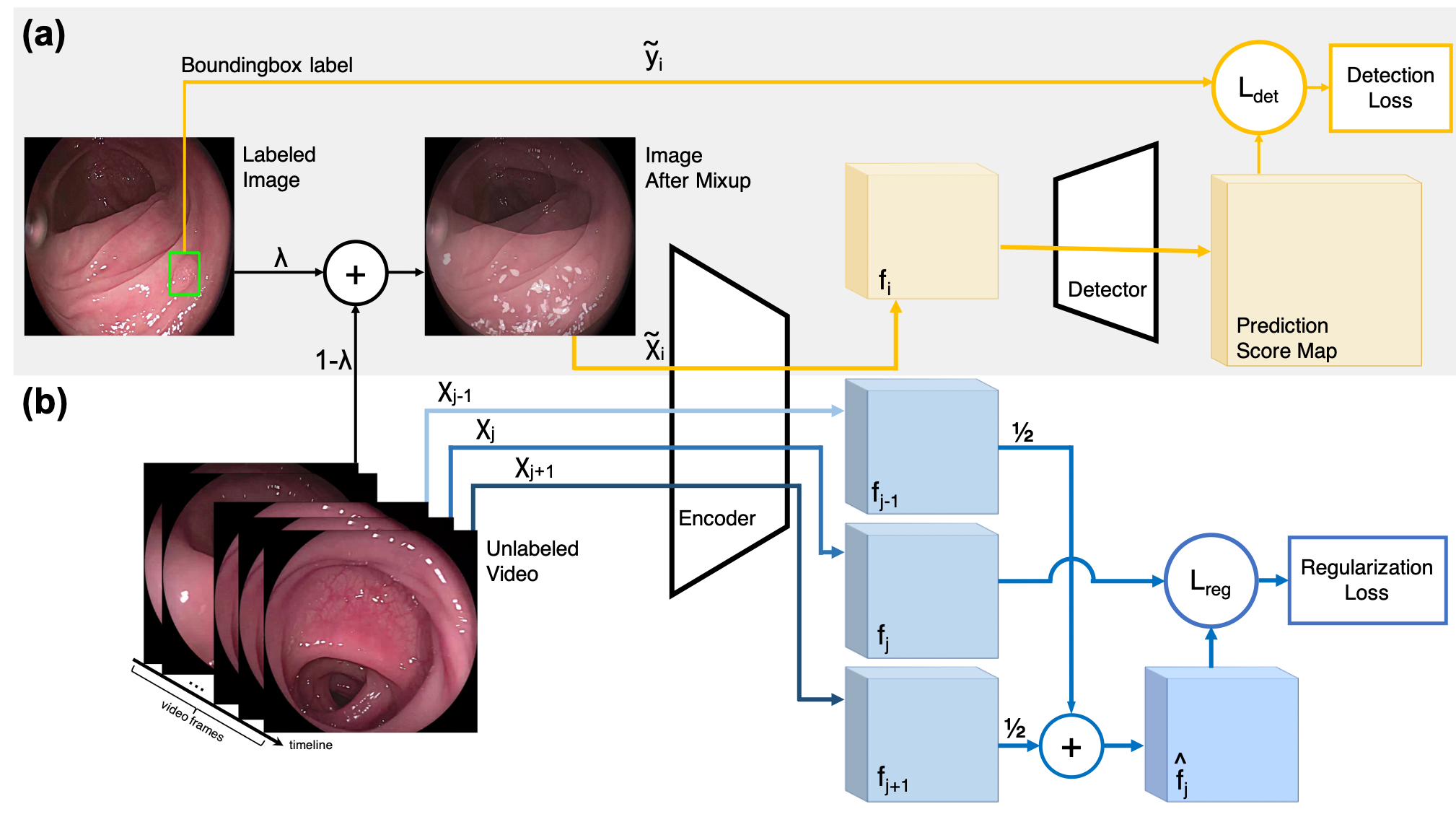}
	\caption{The flowchart of our Ivy-Net.
		(a) We create the virtual input $ \tilde{x_{i}} $ by mixing up the input image with a negative video frame at the pixel level,
		i.e., sum by random weight $ \lambda $,
		without changing its bounding box label $ \tilde{y_{i}} $ (Section~\ref{sec.mixup}).
		(b) $ f_{j - 1}, f_{j}, f_{j + 1} $ are features of adjacent video frames $ x_{j - 1}, x_{j}, x_{j + 1} $,
		encoded by the Encoder of polyp detection model.
		$ \hat{f_{j}} $ is estimation of $ f_{j} $, calculated by $ f_{j - 1} $ and $ f_{j + 1} $,
		e.g. arithmetic mean here.
		Regularization Loss is calculated from $ \hat{f_{j}} $ and $ f_{j} $ by the loss function $ L_{reg} $.} 
	\label{fig.framework}
\end{figure}

\section{The Proposed Approach} \label{sec.approach}
In this paper, we propose an \textit{\textbf{I}mage-\textbf{v}ideo-joint pol\textbf{y}p detection network} (Ivy-Net) to address the domain gap between colonoscopy images from historical medical reports and real-time videos.
By leveraging a large number of unlabeled or weakly labeled videos,
our Ivy-Net improves the performance of colonoscopy polyp detection from two aspects.
Firstly, we want to utilize the semantic information in video frames (see Fig.~\ref{fig.framework}-a).
We augment the labeled images,
constructing virtual samples by mixing them up with negative video frames during model training.
Secondly, we exploit the video time series information (see Fig.~\ref{fig.framework}-b).
We use the coherence between adjacent frames to regularize model training.
The following parts of this section will detail each of these two approaches.

Although compared with the two above,
post-processing which combines detection results from neighboring frames is a more obvious approach.
While in this work, we intend to explore the potential of single-frame detector itself instead.
Our approaches can be used together with post-processing techniques if necessary.

\subsection{Mixing up With Negative Frames} \label{sec.mixup}
Mixup, a data augmentation technique, was first introduced by \cite{zhang2018mixup,inoue2018data},
applied on image classification tasks.
According to the definition in \cite{zhang2018mixup}, it constructs virtual samples $ (\tilde{x}, \tilde{y}) $
$$
\tilde{x} = \lambda x_{i} + (1 - \lambda) x_{j}
$$
\begin{equation} \label{eq.mixup_label_cls}
\tilde{y} = \lambda y_{i} + (1 - \lambda) y_{j}
\end{equation}
where $ (x_{i}, y_{i}) $ and $ (x_{j}, y_{j}) $ are two arbitrary samples from training data;
$ x_{i} $ and $ x_{j} $ are input images;
$ y_{i} $ and $ y_{j} $ are one-hot encoded classification labels;
$ \lambda $ is a random variable in the interval $ [0, 1] $,
controlling the mixup ratio.

\subsubsection{Mixup Labels}
Bounding box labels in image detection tasks are hard to encode as in (\ref{eq.mixup_label_cls}).
\cite{zhang2019bag} simply stacks all bounding boxes from two training images equally.
However in our experiments, the detection precision drops with bounding boxes stacked. 
Therefore, we choose to mixup positive image samples with negative video frames to avoid this problem.
We use the image's bounding box label as the virtual sample's bounding box directly.
\begin{equation}
  \tilde{y} = y_{image_{i}}
\end{equation}

\subsubsection{Mixup Inputs}
Since the images and video frames all have the same aspect ratio,
we simply resize the video frames to the size of images before mixup.
$$
\tilde{x} = \lambda x_{image_{i}} + (1 - \lambda) x_{frame_{j}}
$$
In our experiments, $ \lambda $ 's distribution is also crucial to the detection precision.
We tested two design choices of it.
The first one follows an asymmetric Beta distribution that gives more weight to labeled images:
\begin{equation}
\lambda \sim Beta(\alpha, \alpha + 1)
\end{equation}
where $ \alpha $ is a constant hyperparameter.
Larger $ \alpha $ will give higher average weight to labeled images in our virtual samples.
We define the second discrete probability distribution as follows,
\begin{equation} \label{eq.lambda_discrete}
\lambda = c X
\end{equation}
$$
Pr(X = 1) = p, X \in \{0, 1\}
$$
where X is a random variable that can be 0 or 1,
with the possibility of $ p $ being $ 1 $;
$ p $ and $ c $ are two constant hyperparameters.

\subsection{Temporal Coherence Regularization of Adjacent Frames} \label{sec.feature}
We assume that at high frame rate sampling (e.g. 60fps in our experiments),
adjacent frames of a video contain very similar semantic information.
Hence they should also have similar features after encoding with neural networks.
This similarity provides a kind of supervision that can be used as regularization during training,
forcing the encoder to be smooth in the time dimension.

With this similarity, we further assume the local linearity in the time dimension.
Formally, with encoder $ G $ and parameters $ \theta $,
we get the feature $ f $ from the input $ x $.
\begin{equation} \label{eq.feature}
f = G(\theta, x)
\end{equation}
$ f_{j - 1}, f_{j}, f_{j + 1} $ are features of arbitrary adjacent frames $ x_{j - 1}, x_{j}, x_{j + 1} $.
We define the estimation of $ f_{j} $ as the arithmetic mean of $ f_{j - 1} $ and $ f_{j + 1} $.
$$
\hat{f_{j}} = \frac{f_{j - 1} + f_{j + 1}}{2}
$$
With the local linearity assumption, $ \hat{f_{j}} $ should approach $ f_{j} $ when the frame rate approaches infinity.
We use the cosine distance to measure the similarity of $ f_{1} $ and $ \hat{f_{1}} $.
We calculate the regularization loss $ L_{reg} $ at each point of the feature map,
and then take the average over width and height dimensions.
$$
L_{reg}(\theta, x_{j - 1}, x_{j}, x_{j + 1}) = 1 - cos(f_{j}, \hat{f_{j}}) = 1 - \frac{f_{j} \cdot \hat{f_{j}}}{\| f_{j} \|\| \hat{f_{j}} \|}
$$
The overall regularized detection loss $ L_{reg\_det} $ is the weighted sum of the detection loss $ L_{det} $ and the regularization loss $ L_{reg} $,
$$
L_{reg\_det}(\theta, x_{i}, y_{i}, x_{j - 1}, x_{j}, x_{j + 1}) = L_{det}(\theta, x_{i}, y_{i}) + \gamma L_{reg}(\theta, x_{j - 1}, x_{j}, x_{j + 1})
$$
where $ (x_{i}, y_{i}) $ is an arbitrary labeled image training sample,
and the weight $ \gamma $ is a hyperparameter.

\section{Experiments} \label{sec.experiments}
We will first introduce our datasets and evaluation metrics to further clarify our newly proposed image-video-joint colonoscopy polyp detection task.
Then we will introduce our \textbf{I}mage-\textbf{v}ideo-joint pol\textbf{y}p detection network (Ivy-Net) and training settings.
Finally, we will evaluate each of our approaches respectively.

\begin{figure} [t]
	\includegraphics[width=\textwidth]{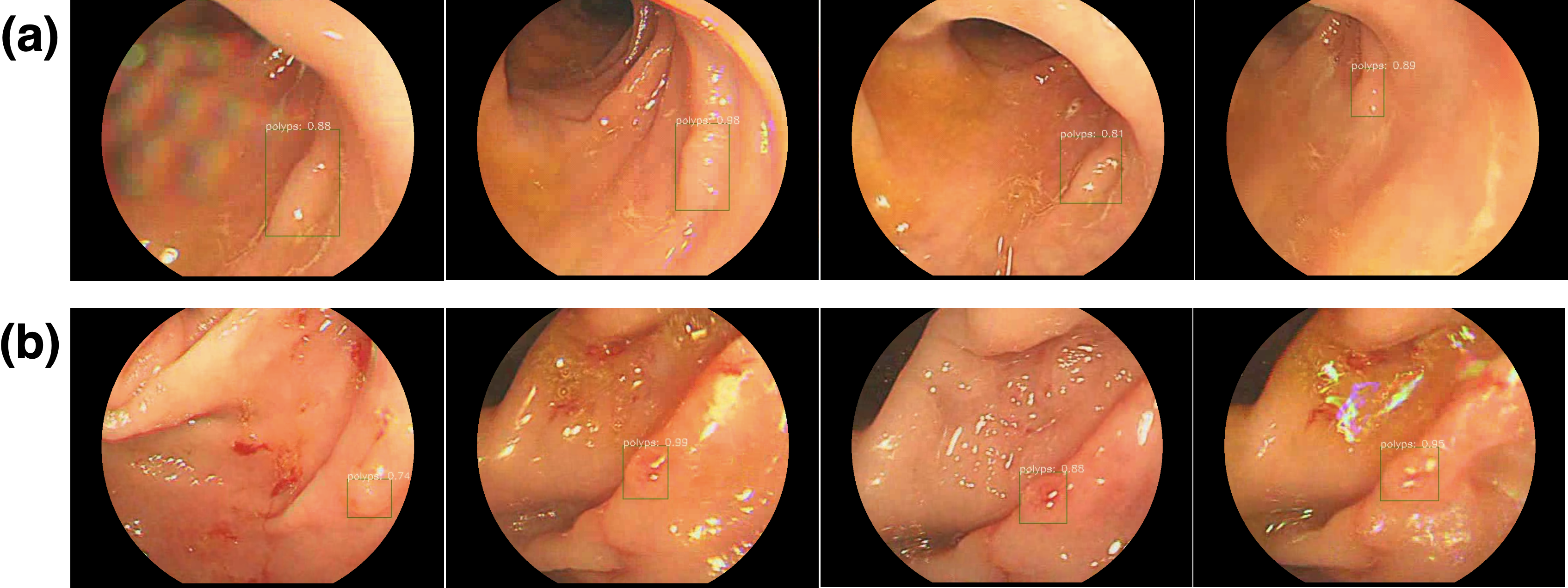}
	\caption{Demo Detection Results on Collected Colonoscopy Dataset.
	(a) and (b) are video frames of two polyps.} 
	\label{fig.demo}
\end{figure}

\subsection{Datasets and Evaluation Metrics}

\subsubsection{Datasets} \label{sec.datasets}
Our colonoscopy dataset contains 3056 images from historical medical reports of 889 positive patients (diagnosed as colon polyp) and 7.5-hour videos (60fps) of 69 patients (28 positive) recorded at *** Hospital.
By professional endoscopists in our team, all images are carefully annotated with bounding boxes;
all videos are annotated with time slice labels of positive or negative;
5050 of positive video frames (12fps sampled) are annotated with bounding boxes.

\noindent \textbf{Data Collection and Annotation:}
All images were captured by Fujinon 4400/4450 systems (Fujifilm Medical Co., Tokyo, Japan), from Nov. 2016 to Oct. 2017;
we used Ultra Stream HDMI (MAGEWELL Co., Nanjing, China) to capture display signal from Fujinon systems and save as videos, during July 2019.
All images were annotated according to Yamada’s classification for colon polyp in white-light colonoscopy with four types,
though only used as one type, i.e., polyp, in this study.
Huashan Institutional Review Board (HIRB) approved this study.
All images, videos and text reports were deidentified.

\noindent \textbf{Splits:}
We randomly split all 3056 labeled images into Image-train set with 2456 samples and Image-test set with 600 samples.
We use all negative video frames (60fps, around 1.5 million) as our Video-train set.
We use all 5050 labeled positive frames as our Video-test set.

Here are two points to clarify.
Firstly, although Video-test set with 5050 video frames seems larger than our image sets,
it comes from only tens of polyps.
These frames are adjacent and similar.
To avoid potential overfitting, we only use them for testing.
Secondly, the task of detection is different from classification.
Models need to distinguish polyps from the background and point out their exact locations.
Wrong position predictions will be counted as false positives.
Besides, clinical applications focus more on models' sensitivity on positive samples.
Hence, using only positive samples for testing is acceptable.

\subsubsection{Evaluation Metrics:}
We use the average precision (AP) at IoU 0.5 to evaluate the polyp detection performance,
which is widely used in object detection tasks.

\begin{table} [!t]
\centering
	\caption{Average Precisions (IoU = 0.5) of Polyp Detection.
		I for Image-train; V for Video-train.
		$ \lambda $ follows the discrete distribution with $ c = 0.5, p = 0.2 $ in MixupDet.
		$ \gamma = 0.01 $ in TCR.
	}
	\label{tab.det_ap_all}
		\begin{tabular}{|c|c|c|c|}
			\hline
			Method & Training Set & AP on Image-test & AP on Video-test \\
			\hline
			Base \cite{ren_faster_2015,lin_feature_2017,he2016deep} & I & 0.770 & 0.631 \\
			MixupDet & I + V & 0.778 & 0.714 \\ 
			TCR & I + V & 0.774 & 0.654 \\
			MixupDet + TCR & I + V & \textbf{0.791} & \textbf{0.717} \\
			\hline
		\end{tabular}
\end{table}

\subsection{Models and Training Settings}

\noindent \textbf{Models:}
We use Faster R-CNN \cite{ren_faster_2015} with FPN \cite{lin_feature_2017} as our basemodel,
and ImageNet \cite{imagenet_cvpr09} pre-trained ResNet-50 \cite{he2016deep} as its backbone, i.e., $ G $ in (\ref{eq.feature}).
$ f $ in (\ref{eq.feature}) is the top FPN feature map, with 256 channels and a stride of 64 pixels with respect to the input image. 

\noindent \textbf{Training Settings:}
We use momentum stochastic gradient descent (SGD) to train all models, for 26 epochs unless otherwise stated.
We use multi-step learning rate with 1e-2 from epoch 0 to 15; 1e-3 from 16 to 21; 1e-4 from 22 to 25.
We apply random horizontal flip on images and corresponding labels with a probability of 0.5.
We freeze the batch normalization (BN) layers of the backbone during training.
Other hyperparameters are as follows: momentum 0.9, weight-decay 1e-4, batch-size 8 (4 GPUs, 2 on each).

\subsection{Evaluation of Mixup}
Experiment results in Table~\ref{tab.det_ap_all} show the effectiveness of our mixup technique.
By mixing negative video frames into labeled images,
the AP (IoU = 0.5) of polyp detection on Video-test set increases significantly from 0.631 to 0.714.
Our mixup technique shrinks the semantic gap between images and videos.
Moreover, the AP on Image-test set also increases from 0.770 to 0.778,
implying the robustness of our algorithm.
So the performance gain on Video-test set is unlikely to be due to overfitting.

An ablation study in Table~\ref{tab.det_ap_mixup} further validates our algorithm,
showing the criticality of $ \lambda $ 's distribution.
As a data augmentation technique, mixup can control the model not to overfit.
But a too strong mixup may cause underfitting instead.
Random variable $ \lambda $ controls its strength.
When $ \lambda $ follows $ Beta(\alpha, \alpha + 1) $,
the lager $ \alpha $ is, the stronger mixup is.
As in Table~\ref{tab.det_ap_mixup}, when $ \alpha $ rises, the detection performance increases first and then decreases.
When $ \alpha $ equals 0.05, we get the best AP of 0.707 in this distribution.
We further test the discrete distribution of $ \lambda $.
When $ c $ equals 0.5 and $ p $ equals 0.2,
i.e., equally mixup image and video inputs with a possibility of 0.2,
we get an even higher AP of 0.714.
However, discrete distribution has two hyperparameters,
$ c $ and $ p $,
which is hard to tune,
especially when there are other regularization hyperparameters.

\subsection{Evaluation of Temporal Coherence Regularization}
Experiment results in Table~\ref{tab.det_ap_all} also indicate our temporal coherence regularization (TCR) to be effective.
With TCR applied,
the AP (IoU = 0.5) of polyp detection increases from 0.631 to 0.654 on Video-test set,
and from 0.770 to 0.774 on Image-test set.
Improvements on both sets validate that our TCR can mine information from unlabeled videos.
With TCR and our mixup technique combined,
the AP (IoU = 0.5) on Image-test set and Video-test set further increases to 0.791 and 0.717 respectively.

\begin{table} [!t]
\centering
	\caption{Ablation Study on MixupDet.}
	\label{tab.det_ap_mixup}
		\begin{tabular}{|c|c|c|c|}
			\hline
			Method & Distribution of $ \lambda $ & Parameters & AP (IoU = 0.5) on Video-test \\
			\hline
			Base \cite{ren_faster_2015,lin_feature_2017,he2016deep} & - & - & 0.631 \\
			\hline
			\multirow{8}{*}{MixupDet} & \multirow{5}{*}{$ Beta(\alpha, \alpha + 1) $} & $ \alpha = 0.02 $ & 0.681 \\
			&  & $ \alpha = 0.05 $ & \textbf{0.707} \\
			&  & $ \alpha = 0.1 $ & 0.698 \\
			&  & $ \alpha = 0.2 $ & 0.675 \\
			&  & $ \alpha = 0.5 $ & 0.627 \\
			\cline{2-4}
			& \multirow{3}{*}{Discrete} & $ c = 0.1, p = 1.0 $ & 0.650 \\
			&  & $ c = 0.2, p = 1.0 $ & 0.665 \\
			&  & $ c = 0.5, p = 0.2 $ & \textbf{0.714} \\
			\hline
		\end{tabular}
\end{table}

\section{Conclusion}
In this study, we defined a clinically needed image-video-joint colonoscopy polyp detection task, and collected a large-scale dataset from 917 positive patients accordingly, over ten times larger than currently available public datasets.
We also proposed a novel Ivy-Net to alleviate the domain gap between colonoscopy images from historical medical reports and real-time videos.
We further exploited the temporal information of unlabeled videos by using temporal coherence regularization (TCR) on feature-level.
Experiments on our collected dataset demonstrate that our Ivy-Net with TCR achieves the state-of-the-art result, improves the AP (IoU = 0.5) of polyp detection on colonoscopy video significantly from 0.631 to 0.717.
Our network is not limited in colonoscopy polyp detection, and could be employed in any image to video joint training task.
We hope this study will offer the community an opportunity to explore more powerful applications.

%
%
%
\bibliographystyle{splncs04}
\bibliography{zhan_bibliography,yang_bibliography}

\begin{thebibliography}{10}
\providecommand{\url}[1]{\texttt{#1}}
\providecommand{\urlprefix}{URL }
\providecommand{\doi}[1]{https://doi.org/#1}

\bibitem{alagappan2018artificial}
Alagappan, M., Brown, J.R.G., Mori, Y., Berzin, T.M.: Artificial intelligence
  in gastrointestinal endoscopy: The future is almost here. World journal of
  gastrointestinal endoscopy  \textbf{10}(10), ~239 (2018)

\bibitem{de2019training}
de~Almeida~Thomaz, V., Sierra-Franco, C.A., Raposo, A.B.: Training data
  enhancements for robust polyp segmentation in colonoscopy images. In: 2019
  IEEE 32nd International Symposium on Computer-Based Medical Systems (CBMS).
  pp. 192--197. IEEE (2019)

\bibitem{arnold2017global}
Arnold, M., Sierra, M.S., et~al.: Global patterns and trends in colorectal
  cancer incidence and mortality. Gut  \textbf{66}(4),  683--691 (2017)

\bibitem{brenner2018colorectal}
Brenner, H., Chen, C.: The colorectal cancer epidemic: challenges and
  opportunities for primary, secondary and tertiary prevention. British journal
  of cancer  \textbf{119}(7),  785--792 (2018)

\bibitem{imagenet_cvpr09}
Deng, J., Dong, W., et~al.: {ImageNet: A Large-Scale Hierarchical Image
  Database}. In: CVPR (2009)

\bibitem{early2012appropriate}
Early, D.S., Ben-Menachem, T., et~al.: Appropriate use of gi endoscopy.
  Gastrointestinal endoscopy  \textbf{75}(6),  1127--1131 (2012)

\bibitem{he2016deep}
He, K., Zhang, X., Ren, S., Sun, J.: Deep residual learning for image
  recognition. In: CVPR. pp. 770--778 (2016)

\bibitem{inoue2018data}
Inoue, H.: Data augmentation by pairing samples for images classification.
  arXiv preprint arXiv:1801.02929  (2018)

\bibitem{itoh2019stable}
Itoh, H., Roth, H., et~al.: Stable polyp-scene classification via subsampling
  and residual learning from an imbalanced large dataset. Healthcare Technology
  Letters  \textbf{6}(6),  237--242 (2019)

\bibitem{leufkens2012factors}
Leufkens, A., Van~Oijen, M., Vleggaar, F., Siersema, P.: Factors influencing
  the miss rate of polyps in a back-to-back colonoscopy study. Endoscopy
  \textbf{44}(05),  470--475 (2012)

\bibitem{lin_feature_2017}
Lin, T.Y., Dollár, P., et~al.: Feature pyramid networks for object detection.
  In: {CVPR}. vol.~1, p.~4 (2017)

\bibitem{mori2018detecting}
Mori, Y., Kudo, S.e.: Detecting colorectal polyps via machine learning. Nature
  biomedical engineering  \textbf{2}(10),  713--714 (2018)

\bibitem{qadir2019improving}
Qadir, H.A., Balasingham, I., Solhusvik, J., Bergsland, J., Aabakken, L., Shin,
  Y.: Improving automatic polyp detection using cnn by exploiting temporal
  dependency in colonoscopy video. IEEE Journal of Biomedical and Health
  Informatics  (2019)

\bibitem{rees2019artificial}
Rees, C.J., Koo, S.: Artificial intelligence—upping the game in
  gastrointestinal endoscopy? Nature Reviews Gastroenterology \& Hepatology
  pp.~1--2 (2019)

\bibitem{ren_faster_2015}
Ren, S., He, K., Girshick, R., Sun, J.: Faster r-cnn: {Towards} real-time
  object detection with region proposal networks. In: Advances in neural
  information processing systems. pp. 91--99 (2015)

\bibitem{strum2016colorectal}
Strum, W.B.: Colorectal adenomas. New England Journal of Medicine
  \textbf{374}(11),  1065--1075 (2016)

\bibitem{urban2018deep}
Urban, G., Tripathi, P., et~al.: Deep learning localizes and identifies polyps
  in real time with 96\% accuracy in screening colonoscopy. Gastroenterology
  \textbf{155}(4),  1069--1078 (2018)

\bibitem{van2006polyp}
Van~Rijn, J.C., Reitsma, J.B., et~al.: Polyp miss rate determined by tandem
  colonoscopy: a systematic review. American Journal of Gastroenterology
  \textbf{101}(2),  343--350 (2006)

\bibitem{wang2019real}
Wang, P., Berzin, T.M., et~al.: Real-time automatic detection system increases
  colonoscopic polyp and adenoma detection rates: a prospective randomised
  controlled study. Gut  \textbf{68}(10),  1813--1819 (2019)

\bibitem{wang2020effect}
Wang, P., Liu, X., et~al.: Effect of a deep-learning computer-aided detection
  system on adenoma detection during colonoscopy (cade-db trial): a
  double-blind randomised study. The Lancet Gastroenterology \& Hepatology
  (2020)

\bibitem{yu2016integrating}
Yu, L., Chen, H., Dou, Q., Qin, J., Heng, P.A.: Integrating online and offline
  three-dimensional deep learning for automated polyp detection in colonoscopy
  videos. IEEE journal of biomedical and health informatics  \textbf{21}(1),
  65--75 (2016)

\bibitem{zhang2018mixup}
Zhang, H., Cisse, M., Dauphin, Y.N., Lopez-Paz, D.: mixup: Beyond empirical
  risk minimization. In: International Conference on Learning Representations
  (2018)

\bibitem{zhang2018polyp}
Zhang, R., Zheng, Y., et~al.: Polyp detection during colonoscopy using a
  regression-based convolutional neural network with a tracker. Pattern
  recognition  \textbf{83},  209--219 (2018)

\bibitem{zhang2019bag}
Zhang, Z., He, T., et~al.: Bag of freebies for training object detection neural
  networks. arXiv preprint arXiv:1902.04103  (2019)

\end{thebibliography}

\end{document}